# An extrasolar planetary system with three Neptune-mass planets

Christophe Lovis[1], Michel Mayor[1], Francesco Pepe[1], Yann Alibert[2], Willy Benz[2], François Bouchy[3,4], Alexandre C.M. Correia[5], Jacques Laskar[6], Christoph Mordasini[2], Didier Queloz[1], Nuno C. Santos[7,1,8], Stéphane Udry[1], Jean-Loup Bertaux[9] & Jean-Pierre Sivan[10]

[1]Observatoire de Genève, 51 ch. des Maillettes, 1290 Sauverny, Switzerland

[2]Physikalisches Institut der Universität Bern, Sidlerstrasse 5, 3012 Bern, Switzerland

[3]Observatoire de Haute-Provence, 04870 St Michel l'Observatoire, France

[4]Institut d'Astrophysique de Paris, 98bis Bd Arago, 75014 Paris, France

[5]Departamento de Física da Universidade de Aveiro, Campus Universitário de Santiago, 3810-193 Aveiro, Portugal

[6]Astronomie et Systèmes Dynamiques, IMCCE-CNRS UMR 8028, 77 Av. Denfert-Rochereau, 75014 Paris, France

[7]Centro de Astronomia e Astrofísica da Universidade de Lisboa, Observatório Astronómico de Lisboa, Tapada da Ajuda, 1349-018 Lisboa, Portugal

[8]Centro de Geofísica de Évora, Colégio Luis Verney, Rua Romão Ramalho, 59, 7002-554 Évora, Portugal

[9]Service d'Aéronomie du CNRS, BP 3, 91371 Verrières-le-Buisson, France

[10]Laboratoire d'Astrophysique de Marseille, Traverse du Siphon, 13013 Marseille, France

**Over the past two years, the search for low-mass extrasolar planets has led to the detection of seven so-called 'hot Neptunes' or 'super-Earths' around Sun-like stars. These planets have masses 5-20 times larger than the Earth and are mainly found on close-in orbits with periods of 2-15 days. Here we report a system of three Neptune-mass planets with periods of 8.67, 31.6 and 197 days, orbiting the nearby star HD 69830. This star was already known to show an infrared excess possibly caused by an asteroid belt within 1 AU (the Sun-Earth distance). Simulations show that the system is in a dynamically stable configuration. Theoretical calculations favour a mainly rocky composition for both inner planets, while the outer planet probably has a significant gaseous envelope surrounding its rocky/icy core; the outer planet orbits within the habitable zone of this star.**

Since the discovery of the first extrasolar planet around a solar-type star ten years ago[1], new detections have been regularly reported by several teams, bringing the number of known planets to more than 170 today[2,3]. The vast majority of these discoveries have been made using the radial velocity technique, that is, the measurement of tiny radial velocity variations of the central star due to the gravitational pull of orbiting planets. This technique, intrinsically sensitive to massive, Jupiter-like planets, has been continuously improved to reach an accuracy of ~1 m s$^{-1}$, leading to the discovery of planets lighter than Neptune on close-in orbits. The low end of the planetary mass distribution is now accessible to radial velocity surveys, whose ultimate detection limits have yet to be established. The accumulation of high-precision radial velocity measurements allows us to continuously refine the orbital parameters of the planets known at present and often reveals the presence of other bodies in the systems. The 17 multi-planet systems detected to date have been the subject of numerous researches studying their formation, dynamical evolution and long-term stability. They show an impressive diversity in planetary masses, orbital distances and dynamical structure, but they all share the common property of being dominated by one or more gaseous giant planets in the Jupiter-mass range. In this Article we present the first observed multiple planetary system composed only of Neptune-mass objects, orbiting the star HD 69830.

**Properties of HD 69830**

HD 69830 is a nearby star located 12.6 pc away from the Sun towards the southern constellation Puppis. It has spectral type K0V and visual magnitude $V = 5.95$[4], making it just visible to the naked eye. In order to determine its basic physical properties, we performed a spectroscopic analysis based on models of stellar atmospheres[5]. We obtain an effective temperature $T_{eff} = 5385 \pm 20$ K and a metallicity [Fe/H] = -0.05 ± 0.02 (that is, 89% of the solar heavy element concentration). From these parameters, and using the appropriate bolometric correction, we derive a total luminosity of 0.60 ± 0.03 $L_\odot$ (where $L_\odot$ is the solar luminosity). By interpolating within grids of theoretical stellar evolution tracks[6,7], we find a stellar mass of 0.86 ± 0.03 $M_\odot$ (where $M_\odot$ is the solar mass) and an age of ~4-10 Gyr. HD 69830 is therefore an old main-sequence star, slightly less massive than the Sun.

This star has recently been under close scrutiny owing to the detection by the Spitzer Space Telescope of a strong infrared excess relative to the stellar photosphere[8]. It is attributed to emission by small grains of crystalline silicates with a size below ~1 μm. The grains have a temperature of ~400 K and must therefore be located close to the star, most probably within 1 astronomical unit (1 AU). These observations are interpreted as the signature of a massive asteroid belt within 1 AU, in which collisional processes continuously replenish the debris disk. Alternatively, the infrared emission might be caused by an evaporating super-comet recently captured onto a close orbit around HD 69830, although this scenario seems less likely.

**High-precision radial velocities**

We have obtained high-precision radial velocity measurements of HD 69830 during the past two years with the HARPS instrument installed on the European Southern Observatory 3.6-m telescope at La Silla Observatory, Chile. HARPS is a high-resolution ($R$=110,000) cross-dispersed echelle spectrograph designed to achieve the highest possible radial velocity accuracy[9,10]. It has demonstrated a long-term precision of ~1 m s$^{-1}$, thereby becoming the most powerful instrument with which to detect extrasolar planets with the radial velocity technique[11,12,13]. HD 69830 is a member of the high-precision sample of nearby stars that we are following closely in order to detect very-low-mass planets. We have obtained 74 data points spanning about 800 days between October 2003 and January 2006 (see Supplementary Information).

The radial velocities have been derived from the extracted spectra by the usual cross-correlation technique with a stellar template, coupled to high-precision wavelength calibration[14]. To estimate the uncertainties on the data points, we quadratically add the photon noise, the guiding error, the wavelength calibration uncertainty and the estimated stellar oscillation noise[9,15], leading to a global error bar of 0.7-1.5 m s$^{-1}$ per measurement. This does not include other noise sources that are intrinsic to the star, such as activity-related radial velocity jitter caused by cool spots at the stellar surface. However, HD 69830 exhibits low chromospheric activity, as can be seen by measuring the re-emission flux at the center of the Ca II H & K lines. The normalized chromospheric emission index $\log(R'_{HK})$[16], obtained from our HARPS spectra, has an average value of $-4.97$, typical for old, quiet K dwarfs. Moreover, HD 69830 has a low projected rotational velocity: we measure $v \sin i = 1.1$ (+0.5/-1.1) km s$^{-1}$ using a calibration of the width of the cross-correlation function[17] (where $v$ is the stellar equatorial rotational velocity and $i$ is the inclination angle of the stellar rotation axis relative to the line of sight). From these indicators we expect very low radial velocity jitter, probably below 1 m s$^{-1}$.

## Orbital parameters for the three planets

The analysis of the radial velocity data reveals multi-periodic variations with a peak-to-peak amplitude of ~15 m s$^{-1}$. A close inspection of the radial velocity curve on short timescales clearly shows a sinusoidal modulation with a period of ~9 days, although successive maxima do not occur at the same radial velocity value, indicating that a second signal is present with a period of ~30 days. We first performed a two-keplerian fit to the data with starting values close to these two periods. The global r.m.s. dispersion of the residuals amounts to 1.57 m s$^{-1}$ and the reduced $\chi^2$ value is 4.19 (with 11 free parameters), meaning that this solution is not satisfactory. Moreover, the residuals around the fit are clearly not randomly distributed, but instead reveal another long-term periodicity at about 200 days. We therefore perform a three-keplerian fit to the radial velocities, which gives a much better result with a global (weighted) r.m.s. of 0.81 m s$^{-1}$ and a reduced $\chi^2$ value of 1.20 (16 free parameters). In this model, the inner planet has a period of 8.667 days, an eccentricity of 0.10 and a minimum mass of 10.2 $M_\oplus$ (where $M_\oplus$ is the Earth's mass). The second planet has a period of 31.56 days, an eccentricity of 0.13 and a minimum mass of 11.8 $M_\oplus$. Finally, the third planet has an orbital period of 197 days, an eccentricity of 0.07 and a minimum mass of 18.1 $M_\oplus$. The list of all orbital parameters for the system can be found in Table 1, while Figure 1 shows the phase-folded radial velocity curves for the three planets. Figure 2 shows two close-up views of the data and best-fit model as a function of time, together with the whole radial velocity curve after removal of the inner planets, thus revealing the long-term variations due to the third planet.

To check if our solution really gives the best fit to the data, we also explored the parameter space with a genetic algorithm. This technique, now routinely used to analyse radial velocity data, has the ability to find the absolute minimum on the $\chi^2$ surface, avoiding the risk of getting trapped in a local minimum. In our case, the genetic algorithm yields orbital parameters that are undistinguishable from those we have found with the simple least-squares minimization, and confirms that the three-planet solution gives a superior fit compared to the two-planet model. Finally, we checked that the radial velocity variations are not partly due to stellar radial velocity jitter by computing the bisector velocity span of the cross-correlation function, which traces possible line shape variations[18]. The bisector turns out to be stable at the level of 0.81 m s$^{-1}$ and is not correlated with any of the orbital periods, including the 31.6-day period which is close to the rotation period of the star (~35 days, estimated from the activity index[16]). If the signal was stellar in origin, bisector variations would occur with an amplitude similar to the radial velocity variations (~5 m s$^{-1}$ for the second planet), and the bisector signal would vary in phase with the stellar rotation period. None of these signatures are observed in our case. We are thus confident that the radial velocity signal is indeed due to orbiting planetary companions.

## Dynamical stability of the system

The multiple planetary system around HD 69830 is unique in that it is the first reported to be composed only of Neptune-mass objects, at least within a few astronomical units. Indeed, the time span of the observations (~800 days) and the precision of the measurements allow us to exclude any Saturn-mass planet orbiting within ~4 AU. When discovering a new system, the immediate question arising is whether or not it is dynamically stable over Myr to Gyr timescales. At first glance, the low planetary masses and small eccentricities suggest a high probability that the HD 69830 system is indeed stable. To investigate this point more thoroughly, we performed numerical N-body integrations[19] assuming coplanarity of the orbits and two different inclination angles (corresponding to different true planetary masses). For both inclinations $i = 90°$ (edge-on) and $i = 1°$ (pole-on), the system turns out to remain stable over at least 1 Gyr. Whereas long-term stability could be expected in the minimum-mass case ($i = 90°$, see Figure 3a), it is notable that the system survives even if the true planetary masses lie in the Jupiter-mass range.

We also considered the possibility that an asteroid belt might exist within ~1 AU of the star, as suggested by the recent observations of the Spitzer Space Telescope[8]. Given the orbits of the three planets, the obvious question is whether or not there are zones of dynamical stability in the system where small bodies could survive in spite of the perturbing effects of the planets. For a uniform grid in semi-major axis (from 0.07 AU to 1.20 AU) and in eccentricity (from 0 to 0.9), massless particles were numerically integrated over two consecutive 1000-year time intervals together with the three planets in the system. The variation of the mean motion frequency over the two time intervals provides a stability criterion for the particles[20]. The results, shown in Figure 3b, indicate that two regions seem to be stable enough to harbour an asteroid belt: an inner region between 0.3 AU and 0.5 AU, and the outer region beyond 0.8 AU (assuming there is no massive, as-yet undetected, planet further away). The observed high temperature of the emitting grains (~400 K) seems to favour the hypothesis of a debris disk in the inner region, although the present observations may not allow us to clearly decide for one of the two regions. Conversely, the presence of a stable debris disk within ~1 AU of the star can be used to constrain the inclination of the system, as too-massive planets would have ejected all other bodies out of the central regions. Simulations show that no particle survives in the inner region for $i < 3°$, corresponding to planetary masses in the Jupiter-mass range. Although not particularly strong, this constraint shows that dynamical studies are able to deliver valuable information for the characterization of planetary systems.

**Formation and composition of the planets**

The discovery of the HD 69830 system also represents a milestone for the understanding of planet formation. In the so-called core accretion model[21], planetesimals accrete material from the protoplanetary disk, forming first a rocky or icy core with a mass up to 10-15 $M_\oplus$. After that, a runaway gas accretion phase follows, which rapidly leads to the formation of a gaseous giant planet provided gas accretion starts before the evaporation of the disk. Simultaneously, inward migration due to interactions with the disk decreases the orbital semi-major axis. The final fate of the planet will be determined by the subtle balance between accretion, disk evaporation and migration timescales[22,23].

Using the models of ref. 22 we performed a large number of simulations aiming at reproducing the HD 69830 system. We assumed a stellar mass of 0.86 $M_\odot$ and slightly sub-solar metallicity, but did not take into account the possible evaporation of the planets[24,25]. We found that in order to reproduce the mass and semi-major axis of these planets, gas disk masses (between 0.07 AU and 30 AU) of 0.04-0.07 $M_\odot$ and disk lifetimes of 1-3 Myr are required. These values are compatible with the observed ones[26,27]. We also found that to account for planets with masses between 10 $M_\oplus$ and 20 $M_\oplus$ at 0.2 AU, a significant amount of migration had to occur (at least one tenth of the value analytically predicted for laminar disks[28]). The starting positions of the embryos of the inner planets are found to be inside the ice line. Thus, these two bodies have grown by accreting essentially rocky planetesimals and some amount of gas. Obviously, if these planets lose mass owing to evaporation, a larger initial mass would be required to account for the present mass, implying that the embryos formed at larger distances. While the innermost planet would probably still start well within the ice line, the second one might accrete some limited amount of ices. The exact ratio between core/envelope mass depends upon whether or not the disk of solids is depleted close to the central star (due to either the high temperatures in the inner disk, or to the formation process of the planets itself), as depletion in planetesimals triggers gas accretion[29]. Assuming no depletion, the envelope of the inner planets is likely to remain very small, whereas strong depletion could lead to a substantial mass of accreted gas (up to 5 $M_\oplus$).

The embryo of the outermost planet started beyond the ice line and therefore accreted a large amount of ices. Again, the core-to-envelope mass ratio depends on the possible depletion of the disk of solids by the formation of the second embryo. In order for the final mass to remain around 20 $M_\oplus$, the Kelvin-Helmholtz time must be of the order of the disk lifetime (a few Myr), which limits the mass of the planet at this time to about 9-10 $M_\oplus$[23]. Therefore, in the case of strong disk depletion by the second embryo, the final planet probably has a core-to-envelope mass ratio of about unity, while low depletion would lead to a smaller amount of accreted gas. Interestingly, this planet appears to be located near the inner edge of the habitable zone, where liquid water can exist at the surface of rocky/icy bodies[30]. Indeed, the habitable zone is expected to be shifted closer to the star compared to our Solar System owing to the lower luminosity of HD 69830. Although this ~20 $M_\oplus$ planet is probably not telluric, its discovery opens the way to an exciting topic in astronomy: the characterization of low-mass planets in the habitable zone of solar-type stars.

Obviously, the planetary system around HD 69830 deserves thorough theoretical and observational investigations owing to its many interesting properties. It represents an important step forward in the characterization of planetary systems and will certainly help us to understand their huge diversity better.

**Supplementary Information** is linked to the online version of the paper at **www.nature.com/nature**.

**Acknowledgements** The data presented here were obtained with the ESO 3.6-m telescope at La Silla Paranal Observatory, Chile. We thank O. Tamuz for the development of the genetic algorithm code, and J. Couetdic for his help in studying the stability of the asteroid belt. We thank the Swiss National Science Foundation (FNRS), the Centre National de la Recherche Scientifique (France) and the Fundação para Ciência e a Tecnologia (Portugal) for their continuous support.

**Author Information** Reprints and permissions information is available at npg.nature.com/reprintsandpermissions. The authors declare no competing financial interests. Correspondence and requests for materials should be addressed to C.L. (christophe.lovis@obs.unige.ch) or M.M. (michel.mayor@obs.unige.ch).

**Table 1. Orbital and physical parameters of the planets in the HD 69830 system.**

| Parameter | HD 69830 b | HD 69830 c | HD 69830 d |
|---|---|---|---|
| Orbital period (days) | 8.667 ± 0.003 | 31.56 ± 0.04 | 197 ± 3 |
| Time of periastron (BJD) | 2453496.8 ± 0.6 | 2453469.6 ± 2.8 | 2453358 ± 34 |
| Eccentricity | 0.10 ± 0.04 | 0.13 ± 0.06 | 0.07 ± 0.07 |
| Longitude of periastron (°) | 340 ± 26 | 221 ± 35 | 224 ± 61 |
| Velocity semi-amplitude (m s$^{-1}$) | 3.51 ± 0.15 | 2.66 ± 0.16 | 2.20 ± 0.19 |
| Semi-major axis (AU) | 0.0785 | 0.186 | 0.630 |
| Minimum mass ($M_\oplus$) | 10.2 | 11.8 | 18.1 |
| Number of data points | 74 | | |
| O-C residuals (m s$^{-1}$) | 0.81 (overall) / 1.50 (early) / 0.64 (late) | | |
| Reduced $\chi^2$ value | 1.20 | | |

The parameters and their formal 1$\sigma$ error bars are those given by the best three-keplerian fit to the data. We checked that planet-planet interactions do not significantly influence the orbital parameters by also performing N-body fits to the data with various inclination angles. As expected, dynamical interactions are so weak that they can be neglected over the short time span of the data, and we therefore use the three-keplerian fit in this paper. The r.m.s. of the observed minus calculated (O-C) residuals is given separately for early and late measurements to illustrate the higher quality of more recent data points. Note that the radial velocity semi-amplitudes of the three planets are the smallest measured to date, showing the potential of the radial velocity method to detect terrestrial planets close to their parent star. BJD, barycentric Julian date.

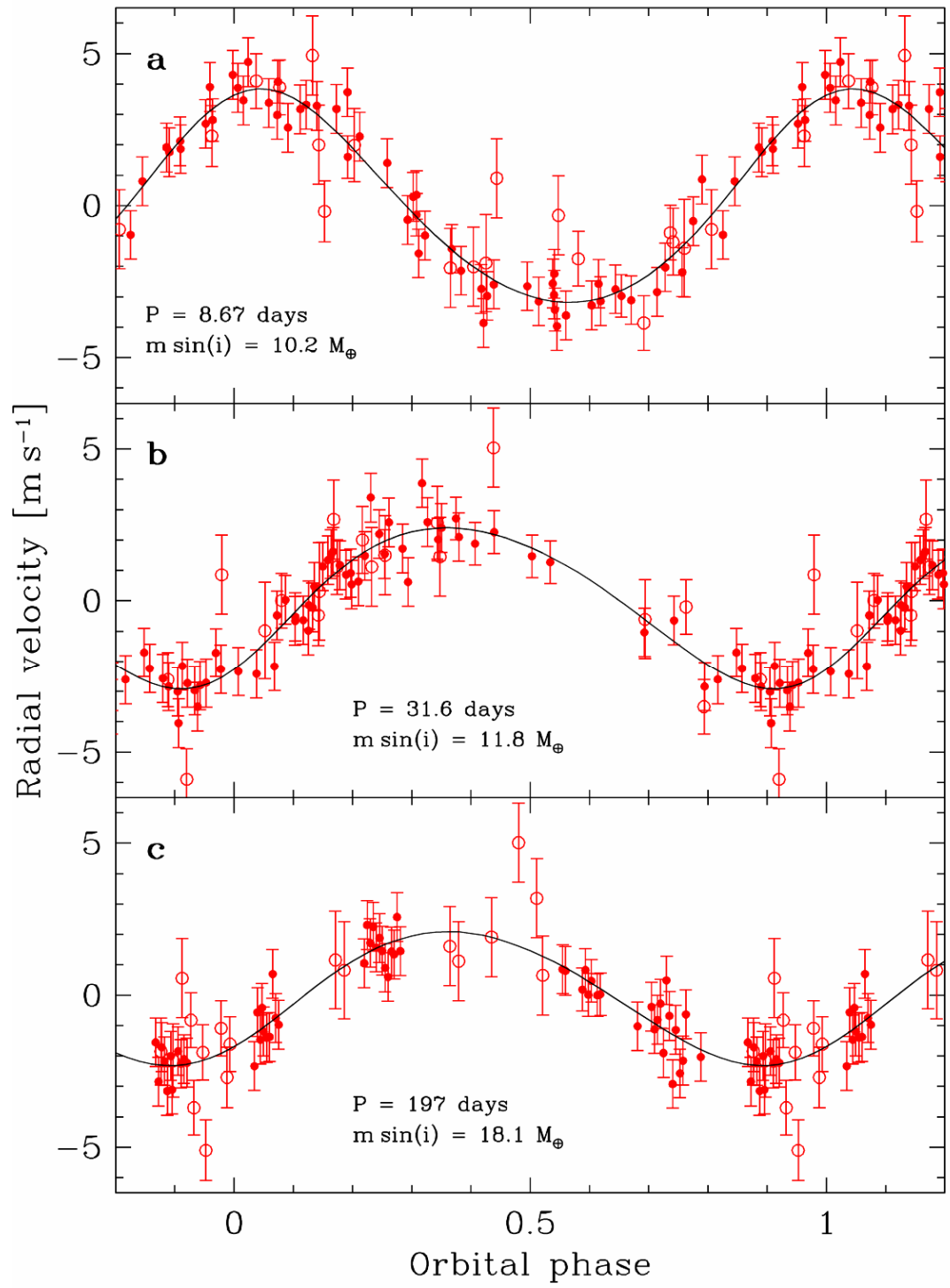

**Figure 1. Phase-folded radial velocity curves for the three planets**. In each case, the contribution of the two other planets has been subtracted. The orbital periods, $P$, are 8.67, 31.6 and 197 days, for the inner (a), intermediate (b) and outer (c) planet, respectively. The radial velocity semi-amplitudes range from 3.5 to 2.2 m s$^{-1}$, corresponding to minimum masses $m$ sin$i$ of 10.2 $M_\oplus$, 11.8 $M_\oplus$ and 18.1 $M_\oplus$ (here $M_\oplus$ is the Earth's mass, $m$ is the actual planetary mass and $i$ is the inclination angle of the system). The integration time was 4 min on average for the first 18 measurements (shown as open circles), and was increased to 15 min for the remaining points (filled circles). The latter measurements are of much higher quality for the following reasons: lower photon noise (from 0.4 to 0.2 m s$^{-1}$), improved guiding accuracy (from ~1.0 to 0.3 m s$^{-1}$), lower wavelength calibration error (from 0.8 to ~0.3 m s$^{-1}$) and better averaging of the stellar p-mode oscillations (which have characteristic periods of a few minutes and individual amplitudes of a few tens of cm s$^{-1}$ that may add up to a few m s$^{-1}$)[9,15]. For the K0 dwarf HD 69830 we estimate that the oscillation noise is between 0.2-0.8 m s$^{-1}$ depending on the exposure time. Combining all these error sources in quadrature, we obtain final 1σ error bars between 0.7 and 1.5 m s$^{-1}$.

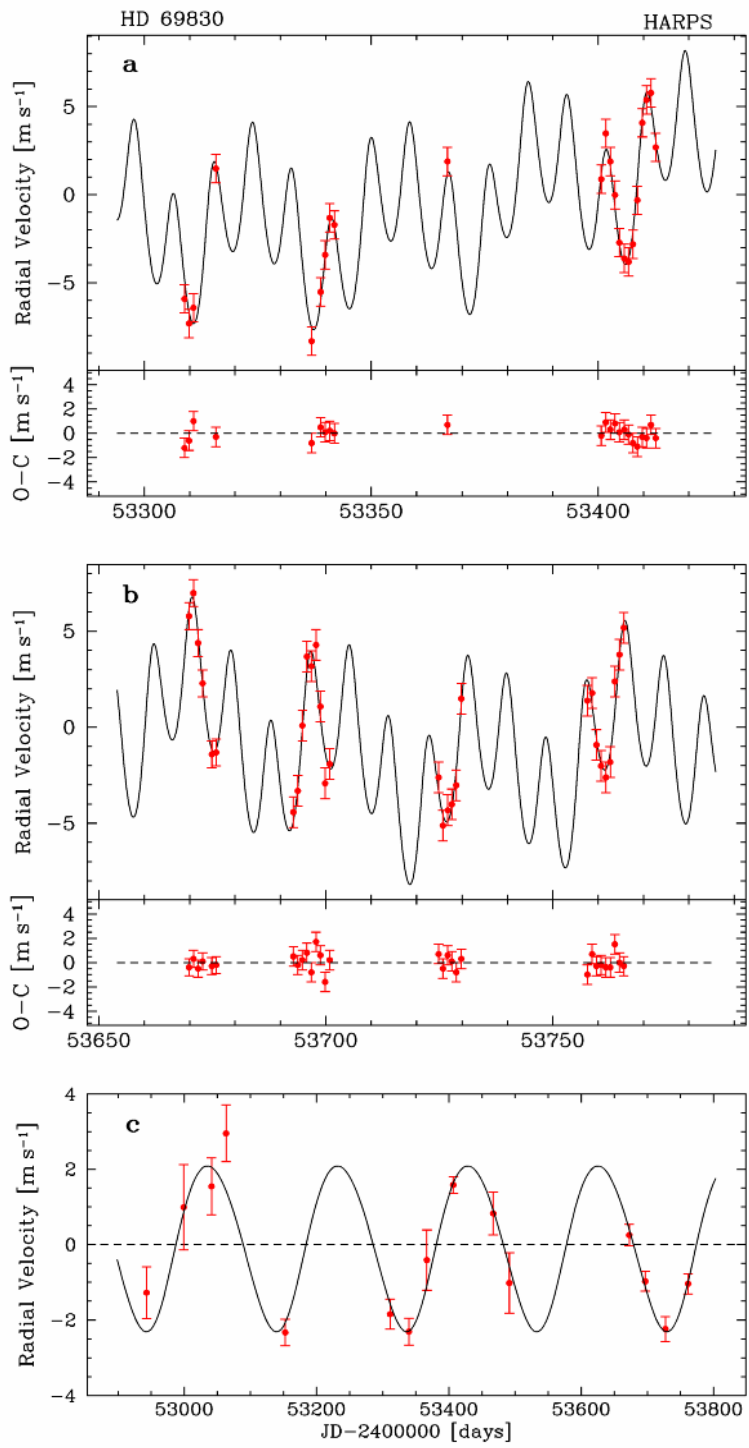

**Figure 2. Radial velocity curve as a function of time. a**, **b**, Close-up views of the data, showing the cumulative signal of three planets. The short-period planet ($P$=8.67 days) shows up as a high-frequency modulation, whereas the intermediate planet ($P$=31.6 days) is revealed through the varying values of successive minima and maxima. The outer planet ($P$=197 days) is not easily seen on these magnified views, but its presence becomes clear when removing the signal of the inner planets and binning the data points (one per observing run), as shown in **c**. Note that only high-quality radial velocity measurements are able to fully resolve this system. The weighted r.m.s. of the residuals around the best-fit model is 0.81 m s$^{-1}$ and becomes as low as 0.64 m s$^{-1}$ when considering only the more recent, higher-quality data points. O-C, observed minus calculated; JD, Julian date. Error bars are 1σ.

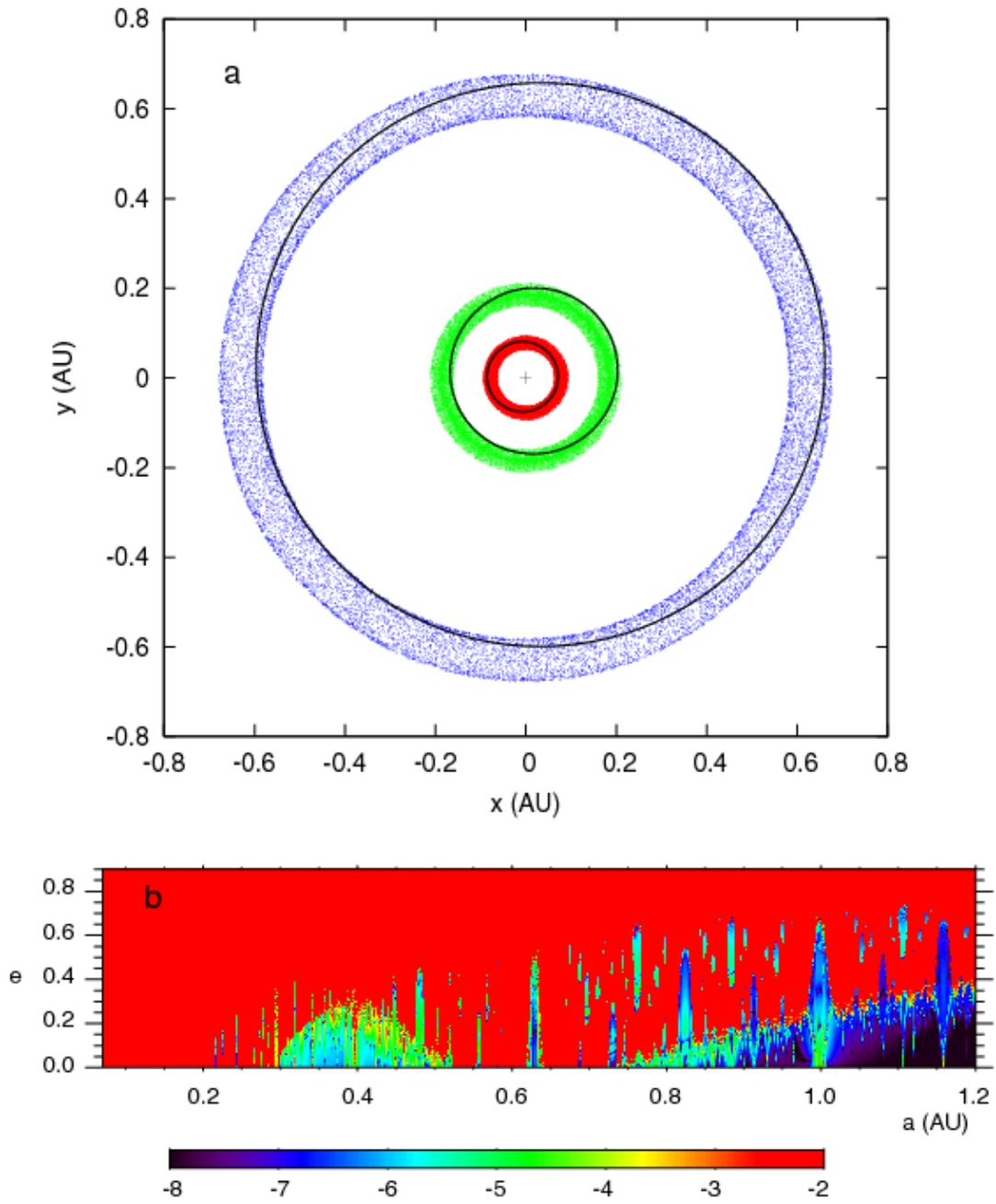

**Figure 3. Dynamical study of the HD 69830 system. a**, Long-term evolution of the orbits of the three planets starting with the orbital solution from Table 1 and inclination equal to 90°. The panel shows a face-on view of the system; $x$ and $y$ are spatial coordinates in a frame centred on the star. Present orbital solutions are traced with solid lines and each dot corresponds to the position of the planet every 50,000 years. The system remained stable for at least 1 Gyr. The semi-major axes are constant and the eccentricities undergo small variations ($0.05 < e_b < 0.20$, $0 < e_c < 0.14$ and $0.069 < e_d < 0.077$, where $e_b$, $e_c$ and $e_d$ denote the eccentricity of the inner, intermediate and outer planet, respectively). The fundamental periods related to the precession of the perihelion are respectively: 5,266 yr, 15,855 yr and 148,000 yr. **b**, Stability analysis for massless particles. For a uniform grid of 0.001 AU in semi-major axis ($a$, from 0.07 to 1.20 AU), and 0.05 in eccentricity ($e$, from 0 to 0.9), massless particles are numerically integrated over two consecutive 1,000-year time intervals together with the three planets in the system. The variation of the mean motion frequency over the two time intervals provides a stability criterion for the particles[20]. The colour grid corresponds to this stability index, red denoting the most unstable orbits with a close encounter with the central star or a planet, while dark blue corresponds to very stable orbits. The particles can survive for an extended time in a region between 0.3 and 0.5 AU, or in the more stable region beyond 0.8 AU. As for the Solar System main asteroid belt, several mean motion resonances with the outermost planet can be identified that would create gaps or accumulation in a potential asteroid belt: 1:2 (~0.40 AU), 2:3 (~0.48 AU), 1:1 (~0.63 AU), 3:2 (~0.82 AU) and 2:1 (~1.00 AU).

## Supplementary information

Supplementary Table 1

This table contains the 74 radial velocity measurements of the star HD 69830 that we have obtained with the HARPS instrument. The format is ASCII text with a tabulation as separation character. The first column contains the epoch of the measurements as Barycentric Julian Date (BJD) minus 2,400,000 (for clarity). The second column gives the radial velocity of the star in km s$^{-1}$ relative to the Solar System barycenter. Finally, the third column gives the uncertainties on the measurements in km s$^{-1}$.

| bjd | radvel | uncertainty |
| --- | ------ | ----------- |
| 52939.87402 | 30.28781 | 0.00115 |
| 52943.87609 | 30.29179 | 0.00123 |
| 52946.86307 | 30.28926 | 0.00127 |
| 52997.80895 | 30.28720 | 0.00159 |
| 53000.71115 | 30.28974 | 0.00157 |
| 53035.77971 | 30.29298 | 0.00129 |
| 53038.69944 | 30.29617 | 0.00134 |
| 53049.62985 | 30.28903 | 0.00128 |
| 53058.63752 | 30.28968 | 0.00133 |
| 53064.60800 | 30.29695 | 0.00132 |
| 53066.62804 | 30.29066 | 0.00132 |
| 53146.50394 | 30.28405 | 0.00090 |
| 53147.46529 | 30.28160 | 0.00094 |
| 53150.45996 | 30.28875 | 0.00087 |
| 53151.45971 | 30.28447 | 0.00097 |
| 53156.51938 | 30.28613 | 0.00088 |
| 53158.47877 | 30.29058 | 0.00104 |
| 53159.47184 | 30.29278 | 0.00089 |
| 53308.84869 | 30.28375 | 0.00081 |
| 53309.85332 | 30.28243 | 0.00081 |
| 53310.83434 | 30.28335 | 0.00079 |
| 53315.78477 | 30.29116 | 0.00076 |
| 53336.87629 | 30.28137 | 0.00081 |
| 53338.87107 | 30.28424 | 0.00081 |
| 53339.87254 | 30.28635 | 0.00081 |
| 53340.87312 | 30.28841 | 0.00078 |
| 53341.87665 | 30.28799 | 0.00077 |
| 53366.80298 | 30.29159 | 0.00078 |
| 53400.70923 | 30.29064 | 0.00084 |
| 53401.69711 | 30.29322 | 0.00079 |
| 53402.70331 | 30.29159 | 0.00077 |
| 53403.73189 | 30.28970 | 0.00077 |
| 53404.67757 | 30.28703 | 0.00077 |
| 53405.78064 | 30.28613 | 0.00077 |
| 53406.72194 | 30.28593 | 0.00077 |
| 53407.68257 | 30.28686 | 0.00080 |
| 53408.64555 | 30.28939 | 0.00080 |
| 53409.74200 | 30.29383 | 0.00079 |
| 53410.66784 | 30.29511 | 0.00078 |
| 53411.65792 | 30.29546 | 0.00078 |
| 53412.69922 | 30.29241 | 0.00078 |
| 53466.61874 | 30.28463 | 0.00076 |
| 53467.52463 | 30.28462 | 0.00076 |
| 53491.48464 | 30.28645 | 0.00077 |
| 53669.86743 | 30.29552 | 0.00072 |
| 53670.82353 | 30.29673 | 0.00073 |
| 53671.84494 | 30.29406 | 0.00072 |
| 53672.85328 | 30.29200 | 0.00072 |
| 53674.87601 | 30.28827 | 0.00073 |
| 53675.84854 | 30.28842 | 0.00072 |
| 53692.84983 | 30.28526 | 0.00078 |
| 53693.82671 | 30.28636 | 0.00076 |
| 53694.84026 | 30.28982 | 0.00077 |
| 53695.82962 | 30.29344 | 0.00077 |
| 53696.81421 | 30.29293 | 0.00076 |
| 53697.84387 | 30.29402 | 0.00078 |
| 53698.80396 | 30.29085 | 0.00077 |
| 53699.83331 | 30.28683 | 0.00076 |
| 53700.86520 | 30.28778 | 0.00076 |
| 53724.85600 | 30.28708 | 0.00076 |
| 53725.80468 | 30.28465 | 0.00076 |
| 53726.84842 | 30.28543 | 0.00077 |
| 53727.76439 | 30.28565 | 0.00076 |
| 53728.74394 | 30.28671 | 0.00076 |
| 53729.86299 | 30.29118 | 0.00077 |
| 53757.64114 | 30.29113 | 0.00077 |
| 53758.69285 | 30.29153 | 0.00076 |
| 53759.64827 | 30.28879 | 0.00076 |
| 53760.65734 | 30.28768 | 0.00076 |
| 53761.71283 | 30.28708 | 0.00077 |
| 53762.66570 | 30.28792 | 0.00077 |
| 53763.70266 | 30.29209 | 0.00077 |
| 53764.73236 | 30.29347 | 0.00077 |
| 53765.66138 | 30.29493 | 0.00077 |